\documentstyle[12pt,epsfig]{article}
\newlength{\defbaselineskip}
\newcommand{\setlinespacing}[1]
           {\setlength{\baselineskip}{#1 \defbaselineskip}}
\newcommand{\onehalfspacing}{\setlength{\baselineskip}%
                           {1.5 \defbaselineskip}}

\newcommand{\be}{\begin{equation}}
\newcommand{\ee}{\end{equation}}
\newcommand{\bea}{\begin{eqnarray}}
\newcommand{\eea}{\end{eqnarray}}

\newcommand{\pr}{{\partial}}

\newcommand{\prm}{\partial_{\mu}}
\newcommand{\prmm}{\partial^{\mu}}

\def\gsim{\lower0.5ex\hbox{$\:\buildrel >\over\sim\:$}}
\def\lsim{\lower0.5ex\hbox{$\:\buildrel <\over\sim\:$}}

\begin{document}



\begin{center}
{\Large\bf Constraining the Randall-Sundrum modulus in the light of recent 
PVLAS data}
\\[12mm]
Debaprasad Maity \footnote{E-mail: tpdm@iacs.res.in}, ~~Sourov Roy
 \footnote{E-mail: tpsr@iacs.res.in}, ~and ~Soumitra SenGupta 
\footnote{E-mail: tpssg@iacs.res.in} \\
{\em Department of Theoretical Physics and Centre for Theoretical Sciences,\\
Indian Association for the Cultivation of Science,\\Kolkata - 700 032, India}\\[10mm]
\end{center}

\begin{abstract} Recent PVLAS data put stringent constraints on the measurement
of birefringence and dichroism of electromagnetic waves travelling in a 
constant and homogeneous magnetic field. There have been theoretical 
predictions in favour of such phenomena when appropriate axion-electromagnetic 
coupling is assumed. Origin of such a coupling can be traced in a low energy 
string action from the requirement of quantum consistency. The resulting 
couplings in such models are an artifact of the compactification of the extra 
dimensions present inevitably in a string scenario. The moduli parameters 
which encode the compact manifold therefore play a crucial role in determining
the axion-photon coupling. In this work we examine the possible bounds on the 
value of compact modulus that emerge from the experimental limits on the 
coupling obtained from the PVLAS data. In particular we focus into the
Randall-Sundrum (RS) type of warped geometry model whose modulus parameter is 
already restricted from the requirement of the resolution of gauge hierarchy 
problem in connection with the mass of the Higgs. We explore the bound on the 
modulus for a wide range of the axion mass for both the  birefringence and 
the dichroism data in PVLAS. We show that the proposed value of the modulus 
in the RS scenario can only be accommodated for axion mass $\gsim$ 0.3 eV.

\end{abstract}

\onehalfspacing

\section{{\large \bf Introduction}      \label{intro}}
Theories with extra spatial dimensions have drawn considerable attention in recent 
times. Various implications of the presence of such dimensions and their observable
consequences are being studied both in the collider and cosmological/astrophysical 
experiments. One of the primary motivations to consider the presence of such extra 
dimension emerge from string theory, which gives us a perturbatively finite quantum 
theory of gravity at the expense of bringing in several extra spatial dimensions.    
Despite many theoretical successes, string theory so far has failed to make any contact
with the observable universe. To establish contact with the present low energy world one 
considers the low energy field theory limit of a string theory. The resulting field 
theory corresponds to the massless sector of a ten dimensional supergravity (SUGRA) 
multiplet coupled to super-Yang Mills. A suitable compactification of the extra dimensions 
then results into an effective low energy supergravity action in four dimensions. It is 
therefore worthwhile to explore some generic observable features of such an action which 
may provide some indirect evidence whether or not a string inspired supergravity model 
is a correct description of our low energy world.  Apart from areas like 
cosmological/astrophysical and high energy collider experiments to look for the signature 
of stringy effects, it is also important to study various purely laboratory based experiments 
which can provide complementary results. In this work we focus into those experiments which 
make use of the conversion of axions or any other low-mass (pseudo)scalar particles into 
photons in the presence of an electromagnetic field. These include the 
Brookhaven-Fermilab-Rutherford-Trieste (BFRT) experiment \cite{bfrt}, the Italian PVLAS 
experiment \cite{pvlas} and several other experiments such as Q \& A \cite{qa}, BMV \cite{bmv} 
etc. which are either already in progress or in the process of being built up. All these 
experiments are expected to produce axions from polarized laser beams, which are allowed 
to propagate in a transverse, constant and homogeneous magnetic field. There has been 
theoretical prediction  of the possibility of modifying the polarization of light propagating 
through transverse magnetic field due to its coupling with the pseudo-scalar axions 
\cite{raffelt}. The resulting birefringence and the dichroism of the vacuum can be tested in 
these experiments, which in turn will constrain the parameter space of a given model involving 
the mass of this light (pseudo)scalar particle and its coupling to the electromagnetic field. 

Other ongoing experiments looking for the (pseudo)scalar-photon mixing
include LIPSS \cite{lipss}, ALPS \cite{alps}, GammeV \cite{gammev} and OSQAR \cite{osqar}.
Axion induced rotation of polarization plane of polarized prompt gamma ray
radiation from gamma ray bursts (GRBs) have been analyzed recently in Ref.\cite{grbs}. 
Several other implications of this axion-two photon coupling have been explored in various 
works \cite{axion-ref}. Such coupling, which is related to the moduli parameters of the extra 
dimension in a string inspired action, therefore can be constrained by experiments.   

The low energy four dimensional effective field theory action of string theory
\cite{gsw,pol,senrev}, namely the supergravity action, contains two massless fields, 
viz., a second rank antisymmetric tensor field (Kalb-Ramond (KR) field \cite{kr}) as well as a 
scalar field called dilaton.  
The interpretation of the KR field strength \cite{pmssg} as a torsion in the background 
spacetime \cite{sabhehl}, inevitably implies the study of electromagnetism in a spacetime with 
torsion. The three form field strength of the two form KR field is identified with
the spacetime torsion and in this context, the gauge $U(1)$ Chern-Simons term that appears 
naturally on account of gauge anomaly-cancellation in the supergravity theory 
plays a crucial role in establishing a gauge-invariant coupling of the KR field 
(or, torsion) with the electromagnetic field \cite{pmssg}. Furthermore the dual
of the Kalb-Ramond field strength in four dimensions in a string inspired model 
is the derivative of a scalar field called axion. Thus an axion-electromagnetic coupling arises
naturally in string theory from the requirement of quantum consistency of the underlying 
supergravity theory. Certain physically observable phenomena, especially in the cosmological 
scenario, may result from such KR-electromagnetic coupling \cite{skssgss, ssgas, dmssgss, 
ssgss, dmpmssg,skpmssgas,skpmssgss}. One such phenomenon of particular interest, argued to be 
induced by the axion-electromagnetic coupling \cite{zavattini}-\cite{pdpj}, is a 
frequency-independent cosmic optical rotation of the plane of polarization of a linearly 
polarized synchrotron radiation from high redshift galaxies as well the birefringence effects 
in electromagnetic waves \cite{birch,kendall,cf,cfj,ralston}. 

In this paper we aim to address these issues in a string inspired model and test it against
the PVLAS experiment to probe into the role of extra dimensions in respect to these
phenomena. Some interesting features of KR background in extra dimensional theories have 
already been explored in several works \cite{bmssssg,bmssgprl,dmssg}. As will be shown later 
that an effective axion-photon coupling in four dimensions, is determined by Planck mass($M_p$) 
and the appropriate parameters encoding the nature of compactification of extra dimensions. 
We shall here focus into a specific axion-electromagnetic coupling induced by compactification 
proposed in Randall-Sundrum warped brane world model. Randall-Sundrum model has successfully 
resolved the problem of fine tuning of the Higgs boson mass against large radiative correction 
( originated from the large hierarchy between the electroweak and the Planck scale) without 
bringing in any new scale in the theory. Apart from supersymmetry, Randall-sundrum model perhaps 
is the most successful theory to resolve this longstanding problem which has always been an 
embarrassment for the standard model of elementary particles. We therefore examine the role of 
RS model in a different context namely the laboratory experiment like PVLAS. It should be noted 
here that there are other particles motivated by string theory which can produce observable 
effects similar to that of axion, in experiments such as PVLAS involving the propagation 
of light in external magnetic field \cite{other-string-motivated}. Millicharged particles
in the context of RS models could also provide similar effects \cite{millicharged}. Recently,
PVLAS results have been analyzed in the context of a chameleon field whose properties 
depend on the environment \cite{chameleon}.    

The plan of the paper is as follows: 
we begin with a brief description of the axion-photon coupling which emerges naturally in a string
inspired model. We shall then consider the Randall-Sundrum warped braneworld model and
explain the nature of the resulting axion-photon interaction. This will be followed by an analysis 
of the optical rotation and the birefringence that result from such interaction. We shall then 
compare our result against the experimental findings of the PVLAS experiments. Finally we shall 
conclude with the possible future directions of our work.

\section{Axion-photon coupling in a string inspired model} 
The string theoretic low energy effective 
action of Einstein-Kalb-Ramond-electromagnetic system is,
\be
S_d  = \int d^d x \sqrt{-G} \left[ M^{d-2} {\cal R} - \frac 1 {12} {\bar H}_{ABC}
{\bar H}^{ABC} - \frac 1 4 F_{AB} F^{AB}\right]
\label{main}
\ee
where, 
\be
{\bar H}_{MNP} = \pr_{[M}B_{NP]} + \frac 1 {M^{d/2 -1}} A_{[M}F_{NP]}.  
\ee
The $U(1)$ gauge field strength corresponding to $A_C$ is given as, $F_{CD} = \pr_{[C}A_{D]}$.
The three form field strength corresponding to the second rank antisymmetric Kalb-Ramond field 
$B_{MN}$ in general is given as  $H_{MNP} = \pr_{[M}B_{NP]}$. However the requirement of 
$U(1)$ gauge anomaly cancellation leads us to redefine $H_{MNP}$ by ${\bar H}_{MNP}$ with the 
appropriate electromagnetic Chern-Simon term as described in Eq.(2) above. ${\cal R}$ is the 
d-dimensional scalar curvature, $M$ is the d-dimensional Planck constant and $\sqrt{-G}$ is 
the determinant of the $d$-dimensional spacetime metric.  

In order to get an effective four dimensional action from the $d$-dimensional action , we need
to compactify the extra $d-4$ dimensions. Keeping the Randall-Sundrum model in mind, let us 
consider the cases of one extra dimension only. After compactification the effective action in 
four dimensions becomes
\be
S_4  = \int d^4 x \sqrt{-g} \left[M_p^2 R - \frac 1 {12} {\bar H}_{\mu\nu\rho}
{\bar H}^{\mu\nu\rho} - \frac 1 4 F_{\mu\nu} F^{\mu\nu}\right]
\ee 
where, 
\be \label{H}
{\bar H}_{\mu\nu\rho} = \pr_{[\mu}B_{\nu\rho]} + \frac \beta {M_p} A_{[\mu}F_{\nu\rho]}.
\ee
In this case the parameter $\beta$ is determined by the geometry of extra dimension and its moduli 
and the compactification scale. $M_p$ is the four dimensional Planck mass and $\sqrt{-g}$ is
the determinant of the $4$-dimensional spacetime metric. The equations of motion 
of the Kalb-Ramond (KR) and electromagnetic fields, obtained from the above action are,

\be \label{HF}
D_{\mu} {\bar H}^{\mu\nu\rho} = 0 ~~;~~ D_{\mu} F^{\mu\nu} = \frac {\beta}{M_p}
{\bar H}^{\nu\rho\gamma} F_{\rho\gamma}.
\ee
The corresponding Bianchi identities are
\be \label{bian}
D_{\mu} {{\tilde F}^{\mu\nu}} = 0~~;~~\epsilon^{\mu\nu\gamma\delta}\pr_{\mu}H_{\nu\gamma\delta} = 0.
\ee
In the above set of equations,  $D_{\mu}$ is the covariant derivative and the dual 
${{\tilde F}^{\mu\nu}} = {1 \over 2} \epsilon^{\mu\nu\gamma\delta} F_{\gamma\delta}$.
Now, in four dimensions the third rank KR field strength tensor can be written in terms
of a massless scalar field called axion through the duality relation,
\be
{\bar H}^{\mu\nu\rho} = \epsilon^{\mu\nu\rho\delta} \pr_{\delta} a.
\ee
In terms of massless axion field $a$, Eqs.(\ref{HF}) and (\ref{bian}) yield
\bea \label{axion-photon}
\Box a = \frac 1 2 F^{\mu\nu} {{\tilde F}_{\mu\nu}}, \\
D_{\mu}F^{\mu\nu} = -\frac {2 \beta}{M_p} \pr_{\mu}a {{\tilde F}^{\mu\nu}}.
\eea
So, from the string inspired low energy effective action, we have an axion-photon coupling 
which depends on the moduli parameters of the compact space and the four dimensional Planck 
mass. We reiterate that our main objective in this paper is  to explore the observable consequences 
of this coupling in recent PVLAS experiments.

So far in this model the axions are taken as massless. However, various stringy 
perturbative/non-perturbative corrections may produce mass for the axion. Without entering 
into the details of its origin, we consider here the mass of the axion as a free parameter
and choose its value in the range which can be probed by the experiments under
consideration. In that case the string inspired effective 4-dimensional low energy action 
becomes,
\be
S = \int \sqrt{-g} d^4x [ M_p^2 R - \frac1 2 ( \prm a \prmm a - 
m_a^2 a^2) - \frac{\beta}{2M_p} a F_{\mu\nu} {{\tilde F}^{\mu\nu}}
- \frac 1 4 F_{\mu\nu} F^{\mu\nu}]
\label{effective-action}
\ee
where, $M_p$ is the 4-dimensional Planck scale, $m_a$ is the mass of pseudoscalar axion 
field $a$ and $F_{\mu\nu}$ is the total electromagnetic field strength. We now calculate 
the exact form of $\beta$ for Randall-Sundrum warped geometric model.

\section{Randall-Sundrum brane world model and its effect on the axion-photon
coupling}
Here we consider a specific five dimensional warped geometry model proposed by Randall and 
Sundrum (RS). The minimal version of such a model is described in a five dimensional bulk 
AdS spacetime where the extra coordinate $y$  is compactified on a ${\mathbf S^1/Z_2}$ 
orbifold. We define $y = r_c \phi$, where $r_c$ is the radius of $\mathbf S^1$ and $\phi$ is 
the corresponding angular coordinate. $Z_2$ orbifolding restricts the value of $\phi$ between 
$0$ and $\pi$.  Two branes, viz., the hidden and visible branes, are located at two 
orbifold fixed points $\phi = 0$ and $\phi = \pi$ respectively. The line element of the 
corresponding background
\be
ds^2 ~=~ e^{- 2 \sigma (y)} \eta_{\alpha\beta} dx^\alpha dx^\beta ~+~ r_c^2 d\phi^2
\ee
describes a non-factorizable geometry with an exponential warping over a flat
($\eta_{\alpha\beta}$) four dimensional submanifold. The warp factor is given in
terms of the parameter $\sigma = k r_c \phi$, where $r_c$ is the compactification
radius and $k$ is of the order of the higher dimensional Planck scale $M$. The
four dimensional Planck scale $M_p$ is related to the five dimensional Planck scale 
$M$ as:~ $M_p^2 = \frac{M^3} k \left(1 - e^{- 2 k r_c \pi}\right)$. The exponential 
warp factor causes a suppression of a scalar mass from the Planck scale to TeV scale 
on the visible brane, located at $\phi = \pi$, as
\be
m = m_0 e^{-k r_c\pi}.
\ee
Thus for $k \sim M_p $ and $r_c \sim 1/M_p$
such that $k r_c \sim 11.7$ , the scalar mass on the brane $m \sim $TeV for $m_0 \sim M_p$. 
Therefore the fine tuning problem in connection with the scalar Higgs mass is resolved 
geometrically without introducing any intermediate scale in the theory.

As an extension to this model, we include the second rank antisymmetric KR field in the bulk 
along with gravity. Just as graviton mode, the KR field, being a closed string excitation, 
can enter into the five dimensional bulk spacetime whereas all the standard model fields 
being open string modes are confined on the visible 3-brane. The Randall-Sundrum 
compactification of the free Einstein-Kalb-Ramond Lagrangian has already been studied 
extensively in \cite{bmssgprl,dmssg}. Similar kind of compactification for the bulk $U(1)$ 
gauge field has been studied in \cite{dhr} and also elucidated in \cite{dmssg} in the context 
of cosmic optical activity. Starting from the Einstein-KR action we 
write down the Eq.(\ref{main}) in $d=5$ as follows,
\be \label{main1}
S_5  = \int d^5 x \sqrt{-G} \left[ M^3 {\cal R} - \frac 1 {12} {\bar H}_{ABC}
{\bar H}^{ABC} - \frac 1 4 F_{AB} F^{AB}\right].
\ee
Let us consider the KR field action
in 5-dimension as follows
\begin{equation}
S_H = \frac 1 {12} \int d^5 x \sqrt{-G} H_{M N L}H^{M N L}
\end{equation}
where $\sqrt{-G}=e^{-4 \sigma} r_c$. This action has KR gauge invariance
$\delta B_{MN} = \partial_{[M}\Lambda_{N]}$.
We use the KR gauge fixing condition to set
$ B_{4 \mu}=0$. Therefore the only non vanishing KR field components are $B_{\mu\nu}$ where
$\mu ,\nu$ runs from 0 to 3. 
These components in general are functions of both compact and
non-compact coordinates. One thus gets
\begin{equation}
S_H = \frac 1 {12} \int d^4 x \int d{\phi}  r_c  e^{2 \sigma(\phi)}
\left[\eta^{\mu \alpha} \eta^{\nu \beta} \eta^{\lambda \gamma} H_{\mu \nu
\lambda} H_{\alpha \beta \gamma} - \frac 3 { r^2_c} e^{-2 \sigma(\phi)}
\eta^{\mu \alpha} \eta^{\nu \beta} B_{\mu\nu} \partial^2_\phi B_{\alpha \beta} \right].
\end{equation}
Applying  the Kaluza-Klein decomposition for the Kalb-Ramond field:
\begin{equation}
B_{\mu \nu}(x,\phi) = \sum^{\infty}_{n=0} B^n_{\mu \nu}(x) \chi^n(\phi) \frac 1
{\sqrt{r_c}}
\end{equation}
and demanding that in four dimensions an effective action for $B_{\mu \nu }$ should be of the form
\begin{equation}
S_H = \int d^4 x \sum^{\infty}_{n=0} \left[ \eta^{\mu \alpha} \eta^{\nu \beta}
\eta^{\lambda \gamma} H^n_{\mu \nu \lambda} H^n_{\alpha \beta \gamma} + 3 m^2_n
\eta^{\mu \alpha} \eta^{\nu \beta} B^n_{\mu \nu} B^n_{\alpha \beta}\right]
\end{equation}
where $H^n_{\mu \nu \lambda} = \partial_ {[\mu} B^n_{\nu \lambda]}$  and $\sqrt
{3} m_n $ gives the mass of the nth KK mode of the KR field,
one obtains
\begin{equation}
- \frac 1 {r^2_c} \frac {\partial^2 \chi^n} {\partial \phi^2} = m^2_n \chi^n
e^{2 \sigma}.
\end{equation}
The $\chi^n(\phi)$ field satisfies the orthogonality condition
\begin{equation}
\int e^{2 \sigma(\phi)} \chi^m(\phi) \chi^n(\phi) d \phi = \delta_{m n}.
\end{equation}
Defining $z_n = e^{\sigma(\phi)} m_n/k$, the above equation reduces to
\begin{equation}
\left[ z^2_n \frac {d^2} {d z^2_n} + z_n \frac d {d z_n} + z^2_n \right] \chi^n
= 0.
\end{equation}
This has the solution
\begin{equation}
\chi^n = \frac 1 {N_n}[ J_0(z_n) + \alpha_n Y_0(z_n)].
\end{equation}
The zero mode solution \cite{bmssgprl,dmssg} of $\chi $ 
therefore turns out to be 
\begin{equation}
\chi^0 (\phi) = C_1 \vert \phi \vert + C_2.
\end{equation}
However, the condition of self-adjointness leads to $C_1 = 0$  and leaves the scope of only a constant
solution for $\chi^{0}(\phi)$. Using the normalization condition, one finally obtains
\begin{equation}
\chi^0 = \sqrt {k r_c} e^{-k r_c \pi}.
\end{equation}
This result clearly indicates that the massless mode of the KR field is suppressed by a 
large warp factor on the visible 3-brane.
In a similar way one can express the electromagnetic field in
bulk first by decomposing it into Kaluza-Klain modes
\be
A_{\mu} (x, \phi) ~=~ \frac 1 {\sqrt{r_c}} \sum_{n = 0}^{\infty} A_{\mu}^n (x)
\xi^n (\phi).
\ee
The solution for the massless mode of the $U (1)$ gauge field \cite{dhr} reads as
\be
\xi^0 ~=~ \frac 1 {\sqrt{2 \pi}} .
\ee

Using the zero mode solutions for both fields, the interaction term in Eq.(\ref{main1}) turns 
out to be 
\bea
 S_{int}= \frac1 {M_p^{\frac 3 2}} \int d^4 x \int d{\phi} \left[e^{2 \sigma}r_c
\eta^{\mu\alpha}\eta^{\nu\beta} \eta^{\lambda\gamma}
H_{\mu\nu\lambda}A_{[\alpha}F_{\beta\gamma]}
+\frac 6 {r_c}\eta^{\mu\alpha}\eta^{\nu\beta}(\partial_{\phi}B_{\mu\nu})A_{\beta}
(\partial_{\phi} A_{\alpha})\right]. \eea 
Now using the Kaluza-Klein
decomposition for both the fields described
earlier,one obtains
\bea
 S_{int}&=& \frac 1 {M_p^{\frac 3 2 }} \int d^4x \int d{\phi}
[e^{2 \sigma}
 \frac 1 {\sqrt {r_c}} \sum^{\infty}_{n,m,l=0}\chi^n \xi^m \xi^l \eta^{\mu\alpha}\eta^{\nu\beta}\eta^{\lambda\gamma}
H^n_{\mu\nu\lambda}A^m_{[\alpha}F^l_{\beta\gamma]}\nonumber\\
& &+ \frac 6 {r_c^{\frac 5 2}} \sum^{\infty}_{n,m,l=0}(\partial_{\phi}
\chi^n)\chi^m (\partial_{\phi}\xi^l)\eta^{\mu\alpha}\eta^{\nu\beta}
 B^n_{\mu\nu}A^m_{\alpha}A^l_{\beta}].
\eea
The part of the above action containing the massless modes only is given as \cite{dmssg} 
\begin{equation}
S_{int}= \frac 1 {M_p^{\frac 3 2}} \int d^4x \int d{\phi}\frac {e^{2 \sigma}} {\sqrt
{r_c}}\chi^0(\xi^0)^2 H_{\mu\nu\lambda}A^{[\mu}F^{\nu\lambda]},
\end{equation}
where $\chi^0= \sqrt {k r_c}e^{-k r_c \pi}$ , $\xi^0= \frac 1 {\sqrt{2 \pi}}$ and
$A^{\mu}=\eta^{\mu\nu}A_{\nu}$.\\
The KR-EM part of the 4d effective action (without the curvature term) therefore becomes 
\begin{equation}
S_{eff}= \int d^4x \left[M_p^2 R - \frac 1 {12}
{\bar H}^{\mu\nu\lambda} {\bar H}_{\mu\nu\lambda} - \frac 1 4 F^{\mu\nu}F_{\mu\nu} \right],
\end{equation}
where 
\be \label{H4}
{\bar H}_{\mu\nu\gamma} = H_{\mu\nu\gamma} + \sqrt{\frac k {M_p^3}} e^{k r_c \pi} A_{[\mu}F_{\nu\gamma]}.
\ee

We have thus explicitly determined the KR-Maxwell coupling in the proposed string inspired RS scenario.
If we now compare the above equation Eq.(\ref{H4}) with Eq.(\ref{H}),  
we observe that the parameter $\beta$ in the effective four dimensional axion-photon coupling 
is determined by the RS compactification as,
\be
\beta = \sqrt {\frac k {M_p}} e^{k r_c \pi}.
\label{beta-compact}
\ee 
Thus determining the axion-photon coupling in a RS compactified string inspired model, we now 
explore the theoretical predictions of such coupling on rotation of plane of polarization as well as
birefringence for an electromagnetic wave propagating in a transverse magnetic field.

\section{Optical rotation and birefringence in string inspired model}
In this section we explicitly estimate the rotation angle of the plane of polarization and 
the birefringence due to the axion-photon coupling that has been shown to arise naturally
in a string inspired model.\\ 
Recall the string low energy action,
\be
S = \int \sqrt{-g} d^4x [M_p^2 R - \frac1 2 ( \prm a \prmm a - 
m_a^2 a^2) - \frac{\beta}{2M_p} a F_{\mu\nu} {{\tilde F}^{\mu\nu}}
- \frac 1 4 F_{\mu\nu} F^{\mu\nu}]
\ee

\be
F_{\mu\nu} = F_{\mu\nu}^{ext} + \prm A_{\nu} - \pr_{\nu} A_{\mu}
\ee
where $F_{\mu\nu}^{ext}$ corresponds to the external magnetic field and
$A_{\mu}$ is the vector potential associated with the light wave.

Now, the classical equations of motion for the axion 
and electromagnetic fields obtained from the above action are
\bea
(\Box  + m_a^2)a = - \frac{2 \beta}{M_p} {\bf E}\cdot {\bf B} \\ 
\nabla \cdot {\bf E} = \frac{2 \beta}{M_p} \nabla a \cdot {\bf B}  \nonumber \\
\nabla \times {\bf B} - \dot{{\bf E}} = \frac{2 \beta}{M_p}[{\bf E} \times \nabla a - {\bf B} \dot{a}] \nonumber \\
\nabla\cdot {\bf B} = 0 \nonumber \\
\dot{{\bf B}} + \nabla \times {\bf E} = 0.
\eea
In the presence of an external magnetic field ${\bf B}_0$, the magnetic field 
${\bf B}$ in the above set of equations can be written as ${\bf B} = {\bf B}_{wave} + {\bf B}_0$, where 
${\bf B}_{wave}$ is the magnetic field associated with the electromagnetic wave.  
In our analysis we will use the gauge condition $\nabla \cdot {\bf A} = 0$. 
Taking the propagation direction of the 
electromagnetic wave to be  orthogonal to the external magnetic field
${\bf B}_0$ and specifying \cite{anselm} the condition ${\bf A}_0 = 0$ the coupled axion-photon equations
in the linear order in $\phi$ and ${\bf A}$ turn out to be,
\bea \label{Aphi}
\Box {\bf A} + \frac{2 \beta}{M_p} \frac {\pr a}{\pr t} {\bf B}_0 &=& 0,\\
(\Box + m_a^2) a - \frac{2 \beta}{M_p} \frac {\pr {\bf A}} {\pr t}\cdot {\bf B}_0 &=& 0.
\eea
However, from the above equation it is clear that only the component
of ${\bf A}$ parallel to ${\bf B}_0$ is affected. Thus for the
linearly polarized wave, the orthogonal component of the 
vector potential with respect to ${\bf B}_0$ can be written as
\bea
{\bf A}_{\perp}(t,x) &=& exp[- i(\omega t - \kappa \cdot x)],\\
{\kappa} \cdot {\bf B}_0 &=& 0,
\eea
where $\omega$ and ${\kappa}$ respectively are the energy and
wave number of the initial beam $(|{\kappa}| = \omega)$. 
Our solution ansatz for the parallel component of the
${\bf A}$ with respect to the external magnetic field ${\bf B}_0$
and the axion field is
\bea
{\bf A}_{||} &=& A_0 exp[- i (\omega' t - {\kappa} \cdot x)],\\
a &=& a_0 exp[- i (\omega' t - {\kappa} \cdot x)].
\eea
So, in order to have the consistent solutions  
we solve the corresponding secular equation
\be
(\kappa^2 - \omega'^2)(\kappa^2 + m_a^2 - \omega'^2) - 
\frac {4 \beta^2 {\bf B}_0^2}{M_p^2} \omega'^2 = 0.
\ee
The roots corresponding to $\omega'$ are
\be
\omega'^2 = \kappa^2 + \delta_{\pm}
\ee
where
\be
\delta_{\pm} = \frac 1 2 \left\{ m_a^2 + \frac {4 \beta^2 {\bf B}_0^2}{M_p^2}
\pm \left[ \left(m_a^2 + \frac {4 \beta^2 {\bf B}_0^2}{M_p^2}\right)^2 + 
\frac {16 \kappa^2 \beta^2 {\bf B}_0^2}{M_p^2} \right]^{1/2}\right\}.
\ee
Using the initial boundary conditions,
\be
{\bf A}_{||}(t=0,x=0) = 1~~~;~~~ a(t=0,x=0) = 0
\ee
the solution becomes,
\be
{\bf A}_{||} = A_1 e^{-i (\omega_{+}t - {\kappa} \cdot x)} + A_2 e^{-i(\omega_{-} t - {\kappa} \cdot x)}
\ee
where, the integration constants are
\bea
A_1  = \frac {- \delta_{-} \omega_{+}}{\delta_{+} \omega_{-} - \delta_{-} \omega_{+}} ,\\
A_2 = \frac {\delta_{+} \omega_{-}}{\delta_{+} \omega_{-} - \delta_{-} \omega_{+}}.
\eea
To establish a connection with the experimental set up, we consider the
initial($t = 0$) electromagnetic field to be
linearly polarized and making an angle $\alpha$ with
the external magnetic field ${\bf B}_0$, so that
\be \label{A}
{\bf A}(t=0) = \cos\alpha ~~{\bf i} + \sin \alpha ~~{\bf j}.
\ee
While travelling through the region of external magnetic field, 
the resulting interaction causes the wave solution to have the form after $t = \ell$ as
\be
{\bf A}(t) =  \cos\alpha ~{\bf A}_{||}(t)~~{\bf i} + \sin \alpha~~ exp(- i \omega t)~~{\bf j}.
\ee
Therefore, the amplitude part of ${\bf A}_{||}$ becomes (up to a common 
phase factor with respect to the orthogonal component ${\bf A}_{\perp}$)
\be \label{Apara}
{\bf A}_{||}(\ell) = A_1 ~exp[-i \theta_+] + A_2 ~exp[-i \theta_-],
\ee
where 
\be
\theta_+ = \frac  {\delta_+ \ell} {2 \kappa}~~~;~~~ \theta_- = \frac  {\delta_- \ell} {2 \kappa}.
\ee
So, from the above set of expressions Eq.(\ref{A}) and Eq.(\ref{Apara}), we see that the vector 
potential describes an ellipse with the major axis at an angle
\be \label{rot}
\alpha(\ell) = \alpha + A_1 ~A_2 ~\sin^2\left(\frac{\Delta \theta} 2\right)~ \sin2 \alpha,
\ee
with the external magnetic field ${\bf B}_0$. Here $ \Delta \theta \equiv \theta_+ - \theta_-$. 
Similarly the extra phase difference developed 
due to interaction with the axion field is 
\be \label{ellip}
\Phi = \tan^{-1} \left[\frac {A_1 \sin \theta_+ + A_2 \sin \theta_-}
{A_1 \cos \theta_+ + A_2 \cos \theta_-}\right].
\ee
Now, Eq.(\ref{rot}) yields the expression for the optical rotation of the plane of polarization 
of the electromagnetic wave as $\epsilon = \alpha(\ell) - \alpha$ and 
similarly the Eq.(\ref{ellip}) gives the expression for 
the ellipticity ${\cal{E}}$ as the ratio of the minor to major axis. 
\bea
\epsilon(\ell) &=& A_1 ~A_2 ~\sin^2\left(\frac{\Delta \theta} 2\right)~ 
\sin2 \alpha ,
\label{rotation}
\eea
\bea
{\cal E}(\ell) &=& \frac 1 2 \tan^{-1} \left[\frac {A_1 \sin \theta_+ + A_2 \sin 
\theta_-} {A_1 \cos \theta_+ + A_2 \cos \theta_-}\right].
\label{ellipticity}
\eea
\vskip .2in
These are the two quantities which establish the direct link with the experimental data.
We now proceed to estimate them in the context of PVLAS experiments.

\section{Probing the moduli parameters using Laser experiments}

As discussed in the previous sections, the purely laboratory based experimental
search for ultra-light (pseudo)scalar particles are devised on the basis of 
the prediction that the polarization properties of light, propagating in a 
constant and transverse magnetic field, can be changed because of the 
couplings of these particles with two photons \cite{zavattini}. In these class 
of experiments, it is possible to make accurate measurements on the 
modification of the polarization state of a light beam. In a practical 
experiment a laser beam is reflected back and forth N times between two 
mirrors, in a constant magnetic field of strength ${\bf B}_0$ which is 
orthogonal to the beam direction. If the distance between 
the subsequent reflections is $\ell$ then the total length travelled by the
laser beam in the magnetic field is $L = N\ell$. The laser beam is linearly
polarized to start with and after traversing a distance $L$, which is usually
of the order of a few kilometers, it is possible to measure very small
ellipticity and change in the rotation of the polarization plane. 

The vacuum magnetic birefringence (or in other words the acquired very small
ellipticity of the linearly polarized light) predicted by the QED, is due to
the dispersive effect produced by the virtual electron-positron pair as 
discussed by Heisenberg and Euler \cite{heisenberg}. The ellipticity produced
this way serves as the background event for the experiment looking for 
birefringence or dichroism produced by (pseudo)scalar particle. The QED 
contribution to the ellipticity can be written as

\begin{eqnarray}
{\cal E} = N {{B^2_0 \ell \alpha^2 \omega} \over {15 m_e^4}},
\label{qed}
\end{eqnarray}
where $\alpha$ = 1/137 is the fine-structure constant, $\omega$ is the photon 
energy and $m_e$ the electron mass. Here we have assumed that the polarization 
vector of the initially linearly polarized beam makes an angle 45$^\circ$ with
the direction of the external magnetic field. If we take a laser beam with
a wavelength $\lambda$ = 1550 nm, $B_0$ = 9.5 T and $N\ell$ = 25 km then the 
resulting ellipticity from Eq.(\ref{qed}) is $2 \times 10^{-11}$ rad 
\cite{battesti}. 

The photon splitting effect can also produce an apparent rotation of the
plane of polarization of a linearly polarized light \cite{adler}. However, 
the resulting effect is too small to be observed in the laboratory. On the
other hand, if the coupling of scalar/pseudoscalar with two photons is 
sufficiently large then this effect of photon splitting can be significantly 
enhanced \cite{emidio,emidio-giovannini}.

It is important to note here that any physical mirror appears to be transparent
to axions so that only the photon component of the beam is reflected. This
essentially sets the axion component of the beam back to zero after each 
reflection \cite{raffelt}. The resulting effect of N reflections is that 
${\cal E}(L) = N {\cal E}(\ell)$ where, in general, $N {\cal E}(\ell)$ is 
not equal to ${\cal E}(N\ell)$.

Thus, in order to take into account the effect of N reflections appropriately 
for a multiple-beam-path experiment, we need to multiply the right-hand side 
of Eq.(\ref{rotation}) and Eq.(\ref{ellipticity}) by N (keeping everything else the same) 
and on the left-hand side the length $\ell$ of a single-path is now replaced by 
the total length $L = N \ell$.  

If we now consider the case of extremely small axion masses, which means 
$\theta_+$, $\theta_-$ and $\Delta \theta \ll$ 1, the results in 
Eq.(\ref{rotation}) and Eq.(\ref{ellipticity}) can be expanded and gives us
\bea
\epsilon(L) = N {{B_0^2} \over {16{\tilde M}^2}}\ell^2
\label{smallmass-rot}
\eea
and 
\bea
{\cal E}(L) = N {{(B_0 m_a)^2} \over {48 \kappa {\tilde M}^2}}\ell^3 .
\label{smallmass-ellip}
\eea
Here the effective inverse coupling constant ${\tilde M}$ is defined as
\bea 
{\tilde M} \equiv M_p/{2 \beta}
\label{relation-beta}
\eea 
as can be seen from Eq.(\ref{effective-action}).
These two results in Eq.(\ref{smallmass-rot}) and Eq.(\ref{smallmass-ellip}) 
agree with the corresponding expressions given by Eq.(44) in Ref.\cite{raffelt} 
\footnote{Note that the relation between $\epsilon(L)$ in our paper and 
$\varepsilon(L)$ in Eq.(44) of Ref.\cite{raffelt} is given by $\epsilon(L)
= {\frac 1 2} \varepsilon(L)$.}.  
As we shall see later, these results in the small axion mass limits describe
the behaviours of the rotation and the ellipticity as a function of 
the axion mass and the inverse coupling strength.

In the year 2006 the PVLAS experiment \cite{pvlas} measured a positive 
value for the amplitude of the rotation $\epsilon$ of the polarization plane in
vacuum with $B_0 \approx$ 5 T. The result is (with a 3$\sigma$ uncertainty)
$\epsilon = (3.9\pm 0.5) \times 10^{-12} ~rad/pass$.
However, the new observations reported very recently \cite{pvlasnew}, do not
show the presence of a rotation signal down to the levels of 
\bea
1.2 \times 10^{-8} ~rad {\rm ~at ~a ~magnetic ~field ~strength ~of ~5.5 ~T} 
\label{limit-rot}
\\ \nonumber
1.0 \times 10^{-9} ~rad {\rm ~at ~a ~magnetic ~field ~strength ~of ~2.3 ~T} 
\eea
(at 95$\%$ c.l.) with 45000 passes. In the same experimental environment no 
ellipticity signal has been detected down to 
\bea
1.4 \times 10^{-8} {\rm ~at ~a ~magnetic ~field ~intensity ~of ~2.3 ~T} 
\label{limit-ellip}
\eea
(at 95$\%$ c.l.). 
These new results exclude the particle interpretation of the previous PVLAS 
results and impose bounds on the mass and the inverse coupling constant for 
scalar/pseudoscalar bosons coupled to two photons. It should be noted that
for the same experimental situation, the QED effects induce a ellipticity 
$\sim 1.6 \times 10^{-10}$.

In Fig. 1 we have plotted the exclusion regions in the two-dimensional 
plane spanned by the parameters axion mass ($m_a$) and the effective inverse 
coupling constant (${\tilde M}$) of axion to two photons. The curves have been 
drawn using our Eq.(\ref{rotation}) and Eq.(\ref{ellipticity}) where we have 
appropriately considered the number of passes (N) and taking into 
account the limiting values for the rotation and birefringence mentioned 
in Eq.(\ref{limit-rot}) and Eq.(\ref{limit-ellip}). We have also assumed the 
value of N to be 45000 in the interaction region. 
From this figure it is evident that the bound on the effective inverse coupling 
${\tilde M}$ coming from the absence of rotation is independent of the mass 
of the axion (for $m_a \lsim 10^{-3}~{\rm eV}$). This can be easily seen from 
Eq.(\ref{smallmass-rot}) where in the small axion mass limit the rotation 
is actually independent of the axion mass. The resulting bound is 
${\tilde M} \gsim 3 \times 10^{6}$ GeV. One can translate this bound on $\tilde M$ 
into an upper bound on the moduli parameter $\beta$ using the relation given
in Eq.(\ref{relation-beta}).Thus, $\beta$ is bounded from above as 
$\beta \lsim {1.6 \times 10^{12}}$. Using the relation between $\beta$ and the 
compactification radius $r_c$ as stated in Eq.(\ref{beta-compact}), one gets an
upper bound on $r_c$ given by
\bea 
kr_c \lsim {1 \over \pi}(13 \ln(10)- \ln(6)) \simeq 8.95.
\label{bound-krc}
\eea
As discussed earlier, the required value of $k r_c$ in order to generate the
TeV scale naturally in  the Randall-Sundrum scenario is $k r_c$ = 11.72. This
value of $k r_c$ is in direct conflict with the bound obtained in 
Eq.(\ref{bound-krc}). Thus, we see that the limits on the 
rotation of plane of polarization of a plane polarized light coming from the
PVLAS experiments, puts severe restrictions on the modulus 
of the Randall-Sundrum scenario. This essentially means that it is very 
difficult to address the hierarchy problem in the context of the RS model,
particularly in the region of low axion-mass. 

\begin{figure}
\includegraphics{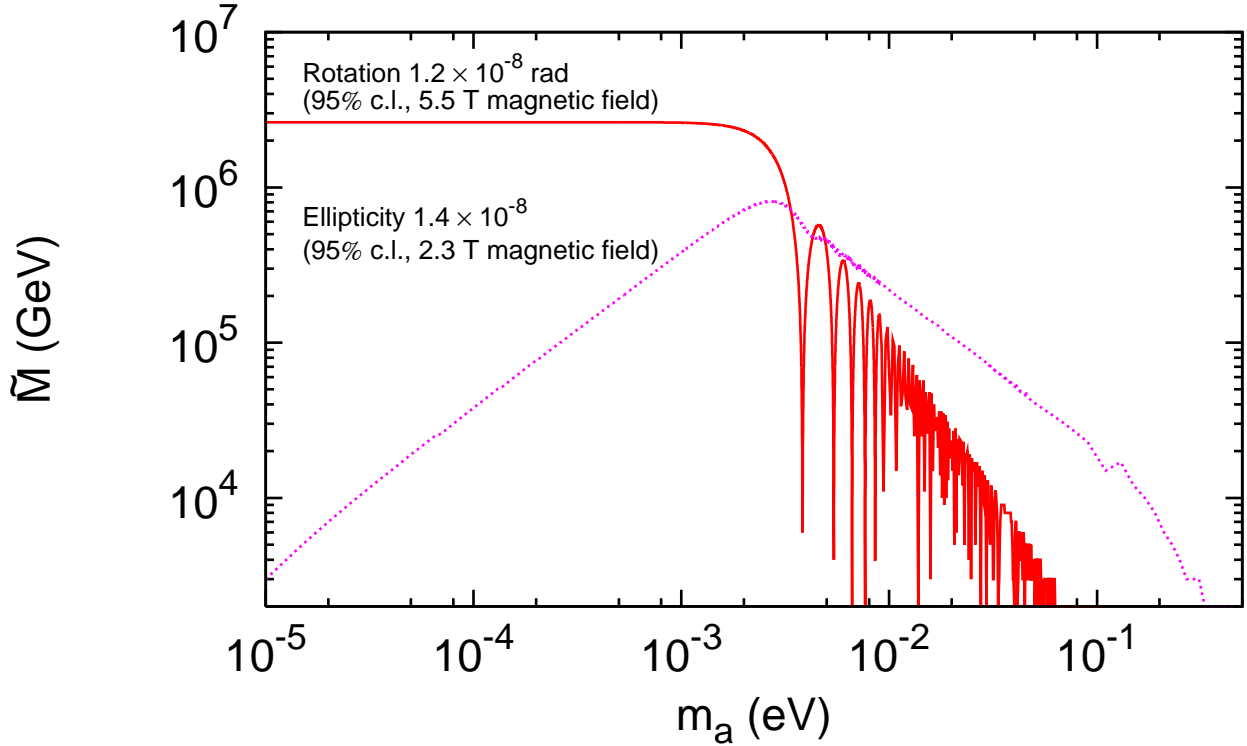}
\caption{\label{fig1} Bounds on mass ($m_a$) and effective inverse coupling 
constant ($\tilde M$) for axion to two photons using the recent PVLAS results
for the rotation and the ellipticity. Area below the solid and the dotted
curves are disallowed from the data.}
\end{figure}

On the other hand, if we consider higher values of the axion mass then the
stronger restrictions on the moduli parameters come from the limits on
the ellipticity measurements as can be seen from Fig.(1). The resulting bound 
on the parameter $\tilde M$ is $\tilde M \gsim 3 \times 10^3$ GeV for an axion
mass $m_a \approx 0.3$ eV. The corresponding limit on the moduli parameters 
appears to be
\bea 
kr_c \lsim {1 \over \pi} (16 \ln(10) - \ln(6)) \simeq 11.15.
\label{new-krc}
\eea
This value of $k r_c$ is more or less in the right ballpark to solve the
hierarchy problem in the Randall-Sundrum scenario. Hence we observe that
though the RS model is disfavoured as a potential candidate to solve the
hierarchy problem in the low axion mass region, it indeed gives the values 
of the parameters in the required range in the region of larger axion mass. 

If we analyze the data for rotation and the ellipticity 
separately then we can see from Fig. 1 that the ellipticity bound allows
the correct value of $k r_c$ (to address the hierarchy problem) also for 
axion mass $m_a \lsim 10^{-5}$ eV. Similarly, the rotation bound gives the
allowed value of $k r_c$ only in the high axion mass region 
($m_a \gsim$ 0.07 eV).

At this stage it is very important to discuss the bounds on the parameter
$\tilde M$ coming from astrophysical considerations. There is a very strong 
constraint $\tilde M \gsim 10^{10}$ GeV (for $m_a < \cal{O}$ (keV) ) 
from the calculation of stellar energy loss \cite{hbstars} of horizontal 
branch stars and from the non-observation of axions in helioscopes such as the 
CERN Axion Solar Telescope (CAST) \cite{cast}. This puts much
severe restrictions on the parameters of Randall-Sundrum model which in turn make such 
models disfavoured in the context of a possible resolution of the hierarchy problem. 
However, it has been shown recently that 
such strong astrophysical constraints can be evaded under certain assumptions 
\cite{redondo}. For example, the axion-two photon vertex can be suppressed at
keV energies due to low scale compositeness of the axion \cite{redondo}. 
One can also consider the case where the temperature and the matter density 
inside the stars control the mass and the coupling of the axion. This way 
axions can acquire an effective mass larger than a few keV which is the 
typical photon energy. This can suppress the axion production inside the 
stellar plasmas and relax astrophysical bounds by several orders of 
magnitude \cite{jaeckel}. Since the stringent astrophysical bounds can be 
evaded, it is very important to look for laboratory experiments where we
have control over all the relevant experimental parameters. This is the reason we have
studied the constraints coming from the PVLAS experiment on the moduli
parameters of the Randall-Sundrum scenario. 

\section{Conclusions}
In this work we have explored the implications of axion-photon coupling in a string inspired 
Randall-Sundrum model where such coupling emerges inevitably from the requirement of quantum 
consistency of the model. Randall-Sundrum model, which is advertised to be a viable alternative 
to supersymmetric theory for offering a possible resolution to the gauge hierarchy problem in 
standard model, confronts some rigorous test in laboratory experiments like PVLAS because of 
such axion-photon coupling. Possibility of finding the signature of warped extra dimensional 
models in controlled laboratory-based experiments is therefore the main motivation of this work.
Our results put severe constraint on the modulus of Randall-Sundrum type
of model. For experiments like optical rotation of the plane of polarization of an electromagnetic
wave, the RS model is disfavoured for axion mass $\lsim 0.07$ eV, whereas for experiments measuring
the ellipticity the value of the modulus reside in the allowed range only for axion mass 
$\lsim 10^{-5}$ eV or $\gsim 0.3$ eV. However on combining both the experimental results the RS model 
is shown to be consistent only for axion mass $\gsim 0.3$ eV. In conclusion RS model, tested against 
PVLAS results( and similar such experiments) puts severe bound on the modulus and the axion mass if 
it has to resolve the hierarchy problem of the standard model. It should be mentioned at this point 
that one can also explain the KR field strength as a torsion in space-time. In that case the role of
additional bulk fields could be important. However, in this work we just extend the original bulk 
gravity model of RS to include the asymmetric torsion part and find its effect in the context of 
PVLAS experiment. We have shown that such a field indeed produces interesting effects and 
determined that to what extent it may explain the experimental data. Analysis presented in this work 
may now be extended for models with more than one extra warped dimensions\cite{dcssg} and also other 
type of compactification scenarios in extra dimensional models. This might lead to a much deeper 
understanding of the role of extra dimension and its possible signature in different laboratory-based 
experiments. 
   
\noindent
{\bf {\Large Acknowledgment}}
\vskip .1in

DM acknowledges the Council of Scientific and Industrial Research, Govt. of India for
providing financial support.

\end{document}